\documentclass[
 reprint, amsmath,amssymb,
 aps, prl,
  groupedaddress
 ]{revtex4-2}
\usepackage{dcolumn}
\usepackage{bm}
\usepackage{mathptmx}
\usepackage{hyperref}
\usepackage{mathrsfs}
\usepackage{amsfonts}
\usepackage[normalem]{ulem}
\usepackage{float}
\usepackage{scalerel}

\usepackage[dvipsnames]{xcolor}
\usepackage{lipsum}  

\usepackage{graphicx}
\usepackage{dcolumn}
\usepackage{bm}
\usepackage{hyperref}
\usepackage{xcolor}

\begin{document}

\preprint{APS/123-QED}


\title{Phase Transitions in Abelian Lattice Gauge Theory:\\ Production and Dissolution of Monopoles and Monopole--Antimonopole Pairs}



\author{Loris Di Cairano}
\email{loris.dicairano@uni.lu}

\author{Matteo Gori}%
\email{matteo.gori@uni.lu}

\author{Matthieu Sarkis}%
\email{matthieu.sarkis@uni.lu}

\author{Alexandre Tkatchenko}
\email{alexandre.tkatchenko@uni.lu}

\affiliation{Department of Physics and Materials Science, University of Luxembourg, L-1511 Luxembourg City, Luxembourg}


\date{\today}

\begin{abstract}
We combine the microcanonical formulation of lattice gauge theories (LGTs) developed by Callaway and the microcanonical inflection point analysis (MIPA) proposed by Bachmann \textit{et al.} to achieve a systematic characterization of phase transitions (PTs) in $U(1)$ lattice electrodynamics. Besides identifying the well-known deconfinement PT (DPT) due to the neutral pair dissolution, which we classify as a first-order PT, we unequivocally detect three higher-order PTs. According to MIPA, we observe two independent third-order PTs in the confined phase; instead, in the deconfined (Coulomb) phase, we observe a dependent third-order PT. For a deeper understanding of the physical meaning of these PTs, we numerically compute the average number density of monopolar and pair defects as a function of energy. Our analysis reveals that DPT is only one of the major mechanisms observable in LGT. The independent third-order PTs are associated, respectively, to the first occurrence of monopolar topological defects and to the production of pairs.

\end{abstract}

\maketitle

Phase transitions (PTs) are collective phenomena that are omnipresent in nature, from protein folding in biology to the formation of starts in astrophysics, or from the superconducting PT in solid state physics to the generation of masses for the weak gauge bosons and fermions known as Anderson-Higgs-Kibble (AHK) mechanism \cite{higgs1966spontaneous,higgs1964broken,englert_broken,higgs1964broken} in quantum field theory (QFT).
In QFT, a deep understanding of PTs would allow to better comprehend how the emergence of order at macroscopic scale is generated from the microscopic scale of elementary quantum components. The origin of this change in scale has been understood in terms of a condensation mechanism of the Nambu-Goldstone boson quanta in the ground state of the system \cite{blasone2011quantum,umezawa1995advanced}. However, such a mechanism holds only for those theories where PTs are triggered by a spontaneous symmetry breaking à la Landau \cite{landau1937theoryI,landau1937theoryII}, that is, when PTs are signaled by non-vanishing values of an order parameter. Although, Landau's mechanism is at the core of many PTs such as, for instance, the AHK mechanism; it is not all-encompassing. Indeed, PTs observed in lattice gauge theories (LGTs), for example, the deconfinement PT (DPT), are not entailed by a symmetry breaking since the gauge symmetry cannot be spontaneously broken due to Mermin-Wagner theorem \cite{mermin1966absence,mermin1967absence}. Also in this case, much work has been done to date in LGTs in order to understand the origin of these PTs and, as shown by Banks et al. \cite{banks1977phase}, 
the DPT is guided by the emergence of topological defects known as magnetic monopoles. First, they interact forming monopole-antimonopole pairs that, in turn, dissociate giving rise to a phase transition \cite{kogut1983lattice,kogut1983lattice,degrand1980topological,barber1985dynamical,barber1984magnetic,grosch1985monopoles,kerler1994phase} similar to the vortex-antivortex pairs dissociation in the Berezinskii-Kosterlitz-Thouless PT \cite{kosterlitz1974critical,kosterlitz1973ordering,berezinskii1971destruction}. 
It is worth noting that the thermodynamic properties of LGTs and consequently the characterization of PTs have so far been performed within the canonical ensemble. The choice of adopting such an ensemble is motivated by the fact that, after Wick rotation \cite{wick1954properties}, the path-integral representation of the vacuum-to-vacuum transition amplitudes \cite{feynman2005space,ar1965quantum} is formally identified with the canonical partition function \cite{peskin1995quantum}.
Despite the success of this approach in providing a physical description of PTs in QFT, an important conceptual inconsistency is inherent in the canonical ensemble. Due to the Lee-Yang theorem \cite{yang1952statistical,lee1952statistical} and Ehrenfest classification \cite{ehrenfest1933phasenumwandlungen}, PTs in (gran)canonical ensemble are signaled by nonanalyticities of thermodynamic observables that can only manifest in the thermodynamic limit.
However, the central goal of LGTs~\cite{wilson1974confinement} is to provide physically meaningful results for the vacuum-expectation values of observables on a finite space-time lattice without employing the continuum limit \cite{callaway1982microcanonical}. Therefore, it would be preferable to develop a method which enables to systematicaly classify PTs for finite-size systems. 

In this Letter, we address this challenge by combining the microcanonical formalism developed by Callaway \cite{callaway1982microcanonical,callaway1983lattice,callaway1985latticesI,callaway1985latticesII,andersen2019quantization,andersen2019quantization,kogut1985further,polonyi1983microcanonical} and the microcanonical inflection point analysis (MIPA) proposed by Bachmann et al. \cite{bachmann2014thermodynamics,bachmann2014novel,schnabel2011microcanonical,qi2018classification,koci2017subphase,sitarachu2020exact,sitarachu2020phase,aierken2020comparison,qi2019influence,koci2015confinement,qi2018classification}.
We remark that MIPA offers a complete classification of PTs that can be applied to systematically identify PTs in the microcanonical ensemble also in finite-size systems, thus dispensing with the need of reaching the thermodynamic limit. MIPA also enables a consistent identification of high-order PTs. For example, MIPA was recently used to discover a new higher-order PT in the 2D Ising model~\cite{sitarachu2022evidence}. Morever, it has also proven to be a powerful method in statistical lattice field theories \cite{bel2020geometrical,pettini2019origin,di2022geometric,pettini2019origin} and for polymers \cite{schnabel2011microcanonical,aierken2020comparison,di2022topological}. Therefore, this fact strongly motivates the application of MIPA in the investigation of the thermodynamics of QFT systems on a lattice.
Specifically, we study one of the fundamental LGTs given by (compact) electrodynamics $U(1)$ on a 4D lattice. Our numerical results allow to classify the DPT as a first-order transition, and two novel (independent) third-order PTs in the confined phase and a (dependent) third-order PT in the deconfined (Coulomb) phase. We show that the physical mechanisms which lead to these transitions are strictly related to peculiar behaviors of topological defects. In order to deeper understand the physical nature of these transitions, a study of the density of isolated monopoles and monopole-antimonopole pairs as functions of energy is provided.
\begin{figure*}[tbh!]
    \includegraphics[scale=0.335]{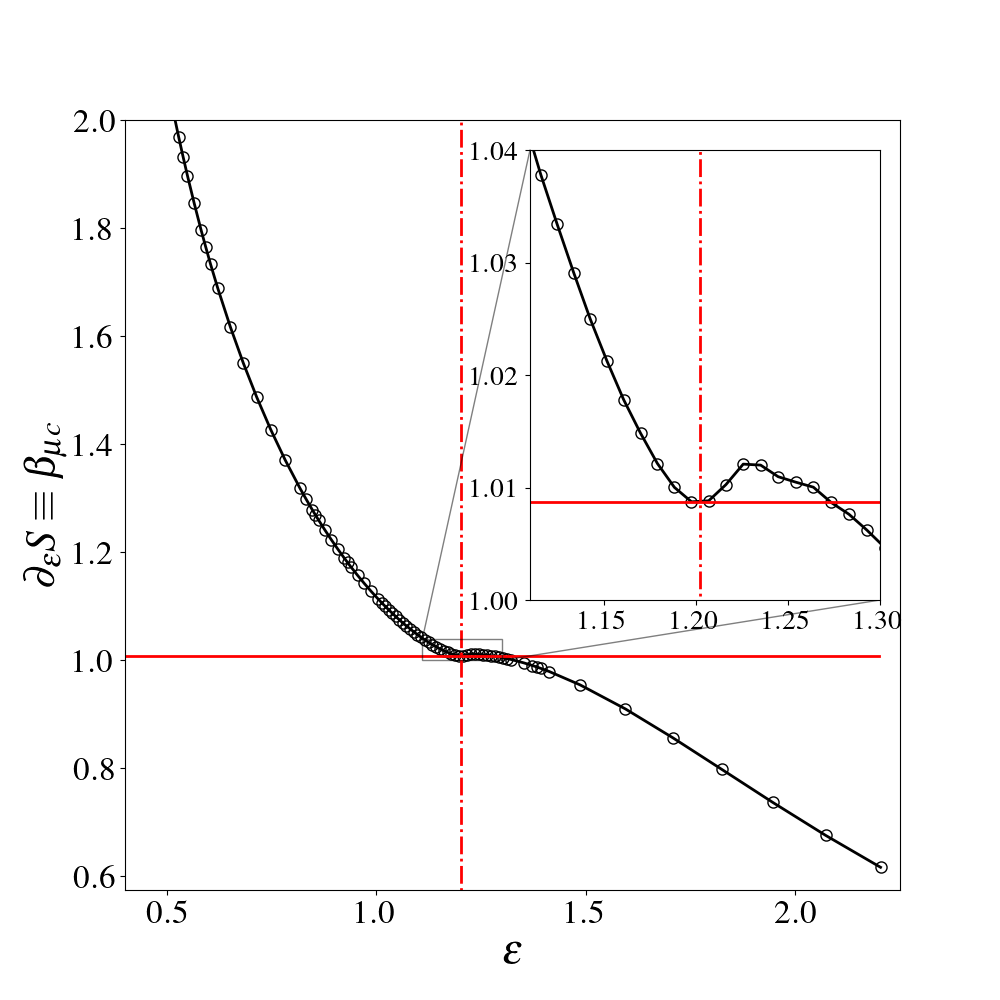}
    \hspace{0.2cm} \includegraphics[scale=0.335]{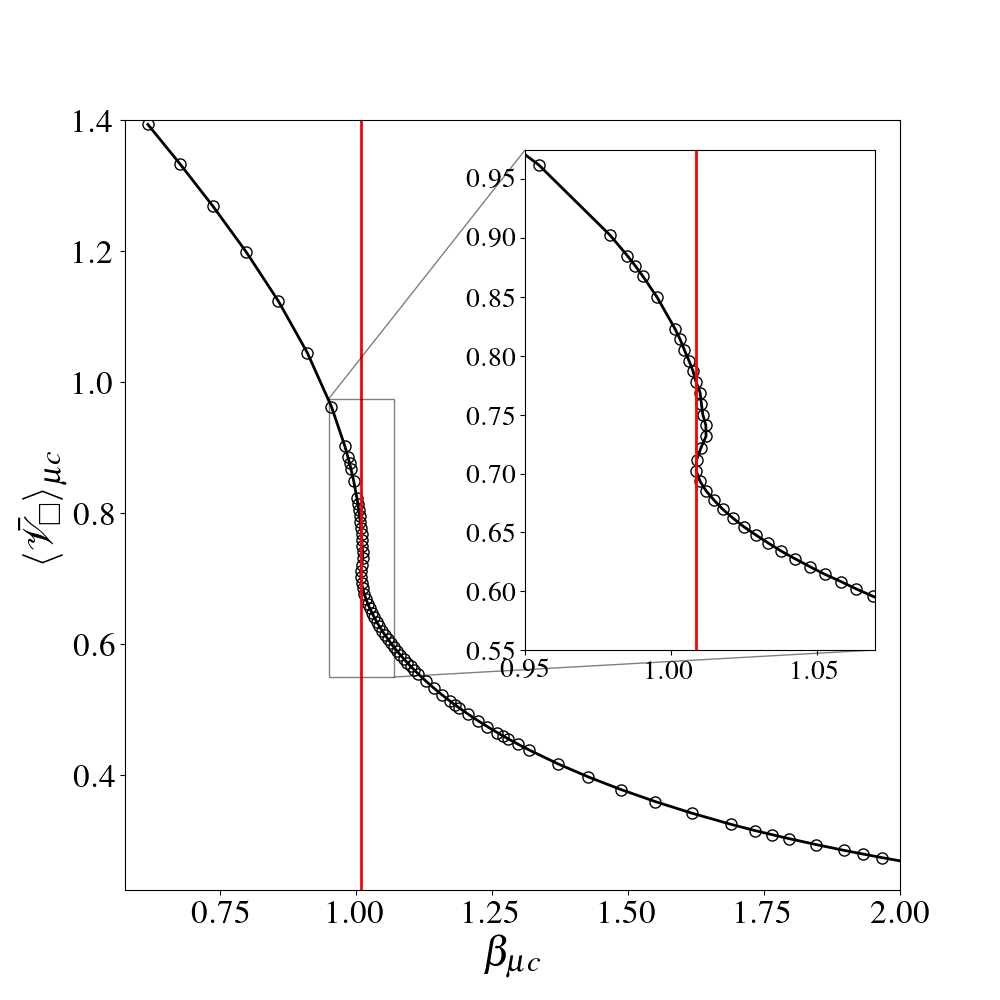}
    \caption{Microcanonical observables for the 4D $U(1)$-LGT with lattice size $N=12^4$. \textbf{Left plot}: First-order derivative of microcanonical entropy, $\partial_{E}S(E)\equiv\beta_{\mu c}$. In the inset, the back-bending region (or S-shape) is shown. The dot-dashed red line passes through the positive-valued minimum of $\beta_{\mu c}$ allowing the identification of the energy value where the first-order transition occurs, i.e., $\epsilon^{ind}_{1}\approx1.197$. The horizontal continuous red line corresponds to $\beta_1=\partial_{\epsilon}S(\epsilon^{ind}_{1})$. \textbf{Right plot}: Average plaquette as a function of inverse temperature, $\langle \bar{\mathscr{V}}_{\square}\rangle_{\mu c}(\beta)$. The inset shows the S-shape of the average plaquette that represents the  fingerprint of a first-order PT in the standard description based on the canonical ensemble. The vertical continuous red line indicates the value of the inverse temperature, $\beta_1=\partial_{\epsilon}S(\epsilon^{ind}_{1})$ corresponding with the deconfinement PT.}
      \label{fig:derS1_average_plaquette}  
\end{figure*}

Let us consider an Abelian gauge field $A_{\mu}(\bm{x})$ discretized over a $d=4$ dimensional Euclidean lattice with $N$ sites per side and a lattice spacing $a$, so that the configuration degrees of freedom are $\phi_{\bm{n},\mu}:=ag_0\,A_{\bm{n},\mu}\in[-\pi/a,\pi/a]$
where $g_{0}$ is the bare coupling constant. Each field variable $\phi_{\bm{n},\mu}$ is related to the link connecting the lattice point $\bm{n}\in[1,N]^{4}$ to its nearest neighbor in the space-time direction $\mu$, i.e. $\bm{n}+\mu$. We employ the gauge-invariant Euclidean Wilson's action \cite{wilson1974confinement}
\begin{equation}\label{def:U1_WilsonAction}
     \mathscr{S}_{W}(\bm{\phi}, \beta):=\beta\sum\limits_{\square}\left[1-\cos(\Theta_{\mu\nu}(\bm{n}))\right] = \beta\sum\limits_{\square} \mathscr{V}_{\square}(\bm{\phi}),
\end{equation}
where the sum over the lattice plaquettes,  $\square=\{\bm{n},\mu,\nu\}$, is
uniquely identified by a site and 
two independent directions on the lattice. Then,
$\Theta_{\mu\nu}(\bm{n}):=\phi_{\bm{n},\mu}+\phi_{n+\mu,\nu}-\phi_{n+\nu,\mu}-\phi_{n,\nu}$ is the plaquette angle, $\mathscr{V}_{\square}$ is the so-called
plaquette potential, and $\beta=g_0^{-2}$ is a parameter that can be interpreted as an effective inverse temperature.
The microcanonical formulation of an Abelian LGT proposed by Callaway \cite{callaway1982microcanonical,callaway1983lattice,callaway1985latticesI} consists in adding a kinetic term $K(\pi)=\sum_{n,\mu}\pi_{n,\mu}^2/2$ to the Wilson's action \eqref{def:U1_WilsonAction} after the introduction of the fictitious momenta, $\pi_{n,\mu}$, conjugated to $\phi_{n,\mu}$. This manipulation does not alter the expectation value of quantum observables, but allows to associate a dynamical system to any LGT whose Hamiltonian function is given by
\begin{equation}\label{def:hamiltonian_LGT}
    \mathscr{H}(\pi,\phi):=K(\bm{\pi})+\beta^{-1} \mathscr{S}_{W}(\bm{\phi}),
\end{equation}
and we naturally identify a microcanonical ensemble. In order to study the thermodynamics, we introduce the entropy function which is the fundamental microcanonical thermodynamic potential that generates all the observables. Hence, the specific entropy function is defined by
\begin{equation}\label{def:microcanonical_entropy}
\begin{split}
    S_{N_{ind}}(E)&:=\dfrac{1}{N_{ind}}\log\Omega_{N{_{ind}}}(E)\\
    &=\dfrac{1}{N_{ind}}\log\int\delta(\mathscr{H}[\pi,\phi]-E)\delta(\mathscr{C}(\pi))~D\pi D\phi\,\,,
\end{split}
\end{equation}
where $\Omega_{N{_{ind}}}(E)$ is the microcanonical partition function and $N_{ind}$ is the number of independent degrees of freedom \cite{callaway1982microcanonical,callaway1983lattice} due to the gauge-fixing constraint $\mathscr{C}(\pi)=0$ (See Supplemental Material (SM) for further details).
Thus, the expectation value of an LGT-observable, $\mathscr{O}[\pi,\phi]$, can be computed by averaging such a function with respect to the microcanonical measure, i.e.:
\begin{equation}\label{def:average_observable_micro}
    \langle\mathscr{O}\rangle_{\mu c}=\Omega_{N_{ind}}^{-1}\int \mathscr{O}[\pi,\phi]\delta(\mathscr{H}[\pi,\phi]-E)\delta(\mathscr{C}(\pi))~D\pi\,D\phi\,.
\end{equation}
where $E$ is the energy of the system which is conserved by the Hamiltonian flow generated by \eqref{def:hamiltonian_LGT} and defined by Hamilton's equations of motion
$d\phi_{n,\mu}/d\tau=\pi_{n,\mu}$ $d\pi_{n,\mu}/d\tau=-\beta^{-1}\partial \mathscr{S}_{\rm W}[\phi]/\partial \phi_{\bm{n},\mu}$. In applications, one numerically solves the Hamilton equations of motion getting a solution $(\pi^{S}_{n,\mu},\phi^{S}_{n,\mu})$; then, assuming in \textit{bona fide} the ergodicity of the trajectory, one recasts Eq.~\eqref{def:average_observable_micro} into a time average:
\begin{equation}\label{def:time_average}
    \langle\mathscr{O}\rangle_{\mu c}\equiv \lim_{T\to\infty}\frac{1}{T}\int_{0}^{T}\mathscr{O}[\pi^{S}_{n,\mu}(\tau),\phi^{S}_{n,\mu}(\tau)]~d\tau\,.
\end{equation}
It should be noticed that the introduction of fictitious momenta leads inevitably to introduce a fictitious time, $\tau$, along which the system evolves. This construction makes the (4+1)D microcanonical formulation equivalent to the stochastic quantization \cite{callaway1984stochastic}. 

\begin{figure}
    \includegraphics[scale=0.335]{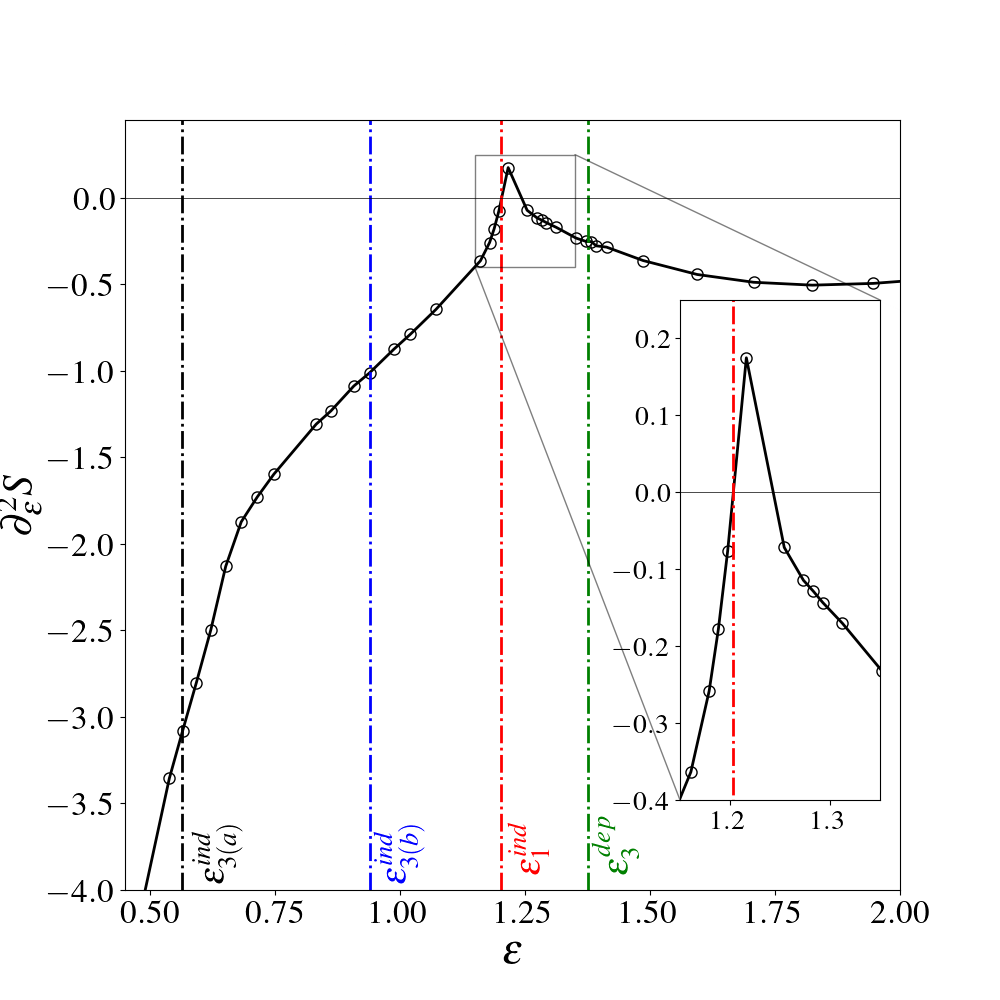}
        \caption{\textbf{Second-order derivative of the microcanonical entropy for the 4D $U(1)$-LGT with lattice size $N=12^4$}. All of the PTs discussed in the main text are represented in the plot by vertical lines. Independent third-order PT, $\epsilon^{ind}_{3\,(a)}$: black. Independent third-order PT, $\epsilon^{ind}_{3\,(b)}$: blue. Independent first-order PT, $\epsilon^{ind}_{1}$: red. Dependent third-order PT, $\epsilon^{dep}_{3}$: green. The panel shows the behavior of the second-order derivative around the deconfinement PT. $\partial_{\epsilon}^{2}S$ vanishes in correspondence of $\epsilon^{ind}_{1}$ with a subsequent maximum as expected in a first-order PT.}
        \label{fig:derS2}
\end{figure}

\begin{figure}
    \includegraphics[scale=0.29]{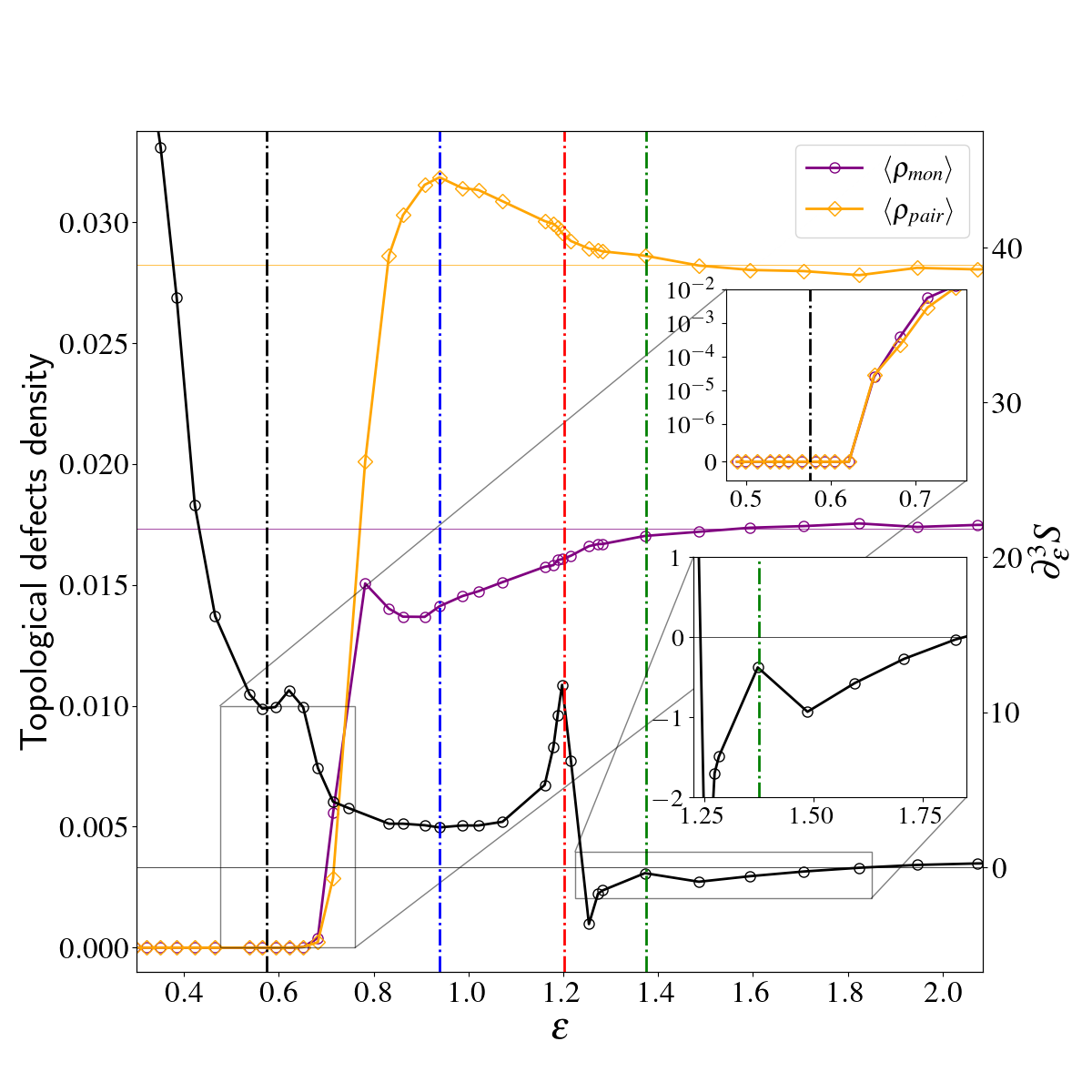}

        \caption{\textbf{Comparison between third-order derivative and density of monopoles for the 4D $U(1)$-LGT with lattice size $N=12^4$}. The black curve represents $\partial_{\epsilon}^{3}S$ whereas the yellow and violet curves are, respectively, the number density of monopole-antimonopole pairs and isolated monopoles. The vertical colored lines indicate the energy value of each PT as indicated in Fig.~\ref{fig:derS2}. Inset on the top: zoom on the energy range where the first appearance of monopoles is observed. Inset on the bottom: zoom on the negative-valued maximum associated to the dependent PT.}
        \label{fig:derS_3_monopoles}
\end{figure}

Having defined a microcanonical description, we can investigate the thermodynamics properties of the $U(1)$-LGT adopting the microcanonical inflection point analysis (MIPA) \cite{schnabel2011microcanonical,bachmann2014thermodynamics, qi2018classification}.
According to MIPA, PTs are split into two classes: independent and dependent. 
Independent PTs represent the major change in the system's phase, whereas a dependent PT can only occur if an independent one already manifested; thereby, they are considered the precursors of a major transition.
An independent PT of even (odd) order $2n$ ($2n-1$) occurs at $E_c$ if $\partial^{2n-1}_{E}S_{N}$ ($\partial^{2n-2}_{E}S_{N}$) admits an inflection point at $E_c$ and $\partial^{2n}_{E}S_{N}$ ($\partial^{2n-1}_{E}S_{N}$) combined with a negative-valued maximum (positive-valued minimum) at $E_c$. 
A dependent PT of even (odd) order $2n$ $(2n + 1)$ is detected at $E_c$, if $\partial_{E}^{2n- 1}S$ ($\partial_{E}^{2n}S$) admits a least-sensitive inflection point at $E_c$ combined with positive-valued minimum (negative-valued maximum) at $\partial_{E}^{2n}S$ ($\partial_{E}^{2n+1}S$)
\cite{qi2018classification}. 
We present here the application of the MIPA to the previously mentioned microcanonical formulation of the 
$U(1)$-LGT. 
In the left panel of Fig.~\ref{fig:derS1_average_plaquette}, the first-order derivative of entropy $\partial_{\epsilon}S$ (coinciding with the inverse of microcanonical temperature $\beta_{\mu c}$ ) as a function of the energy density, $\epsilon=E/N_{ind}$ is presented(see SM for details about such a computation). 
In order to calculate the derivative of microcanonical entropy, we adopted a method introduced by Pearson et al. \cite{pearson1985laplace} allowing to estimate the $n$-order derivative of 
microcanonical entropy at given energy as a function of the time averages \eqref{def:time_average} of the first $n$ positive integer powers of the inverse of the kinetic energy, for instance, $\partial_{E}S(E)= (N_{ind}/2-1)\langle K^{-1}\rangle_{\mu c}\label{def:first_derivative}$ (see SM for higher order derivatives). 
We can easily identify a back-bending region (S-shape) in the inverse of microcanonical temperature $\partial_{\epsilon}S$ entailing a positive-valued minimum of at $\epsilon^{ind}_{1}\approx 1.197$. Such a minimum corresponds to a positive-valued maximum in $\partial^2_{\epsilon}S$ (see Fig.~\ref{fig:derS2}), thus to an independent first-order PT, according to the MIPA.
Such a transition corresponds to the deconfinement phase transition (DPT). 
Indeed, the value of $\beta_{\mu c}$ at the transition point $\epsilon^{ind}_{1}$ corresponds to $\partial_{\epsilon}S(\epsilon^{ind}_{1})\equiv\beta(\epsilon^{ind}_{1})=1/T_1\approx 1.00872$ in unit of $\hbar= k_{B}=1$ which coincides with the transition value of $\beta$ predicted by Monte Carlo simulations  \cite{creutz1979monte,bhanot1981nature,stump1987remarks,jersak1983u,callaway1982microcanonical,callaway1983lattice}. 
It should be stressed that the DPT has sometimes been classified as a second-order PT~\cite{moriarty1982monte,creutz1979monte,callaway1985latticesI,creutz1983monte,bhanot1981nature,degrand1981potential,caldi1983well}, but many evidences, such as the emergence of metastable states, led to conclude that the DPT is of first-order \cite{jersak1983u,sarkar2021study,lautrup1980phase,stump1987remarks,barber1984magnetic,grosch1985monopoles,evertz1985tricritical}. In order to confirm the identification of such a first-order PT with the DPT, we provide the microcanonical computation of the average plaquette, $\langle \bar{\mathscr{V}}_{\square}\rangle_{\mu c} = \langle \mathscr{V}\rangle_{\mu c}/N_{ind} $ as a function of the microcanonical temperature, 
as suggested in Ref. \cite{callaway1982microcanonical}.
As shown in the plot on the right in Fig.~\ref{fig:derS1_average_plaquette}, the average plaquette becomes a multi-valued function that admits an S-shape around $\beta_1\approx 1.00872$ (see the inset); this is a typical signature of a first-order PT.
It has to be noticed that such an S-shape is a direct consequence of the inflection point in $\partial_{\epsilon}S$. In particular, the local minimum associated with the DPT ($\partial_{\varepsilon}S|_{\varepsilon_1}=0$) is the deep origin of the vertical tangent point indicated by the vertical red line on the right panel of Fig.~\ref{fig:derS1_average_plaquette}, i.e.
\begin{equation}
    \lim_{\beta_{\mu c}\rightarrow \beta_1^{+}} \dfrac{\partial \langle \mathscr{\bar{V}}_{\square} \rangle_{\mu c}}{\partial \beta_{\mu c}} = \lim_{\beta_{\mu c}\rightarrow \beta_1^{+}} \dfrac{\partial \langle \mathscr{\bar{V}}_{\square} \rangle_{\mu c}}{\partial \varepsilon} \dfrac{\partial\varepsilon}{\partial \beta_{\mu c}} = -\infty,
\end{equation}
as $\partial_{\varepsilon}\langle\bar{\mathscr{V}}_{\square}\rangle$ is positive and bounded on the considered range of energies (see Fig.~2 of SI)
and $\lim_{\beta_{\mu c}\rightarrow \beta_1^{+}} \partial_{\beta_{\mu c}} \varepsilon =-\infty$.
Following the classification scheme associated with MIPA, we can identify a further third-order independent PT, associated with an inflection point in $\partial^2_{\epsilon}S$ (see the vertical blue line in Fig.~\ref{fig:derS2}) together with a positive-valued minimum in $\partial^{3}_{\epsilon}S$ around $\epsilon^{ind}_{3\,(b)}\approx 0.93$ (see Fig.~\ref{fig:derS_3_monopoles}).
In addition, we have good evidence that encourages the identification of two other third-order PTs; the first is independent, whereas the second is dependent. 
The former (see the vertical black line in Fig.~\ref{fig:derS_3_monopoles}) occurs around $\epsilon^{ind}_{3\,(a)}\approx0.57$ and the latter (see the vertical green line in Fig.~\ref{fig:derS_3_monopoles}) around $\epsilon^{dep}_{3}\approx 1.375$. To understand the physical meaning of the \emph{large} number of PTs identified by MIPA and never observed so far, we analyze and compare the third-order derivative of the microcanonical entropy with
the average number density of topological defects. 
The current knowledge of the role of topological defects during the DPT is summarized in Ref.~\cite{banks1977phase}: at $T<T_1$, there are few monopole loops, and they will be small in spatial extent. 
Increasing $T$ increases the density and size of loops. At $T_1$, the monopole-antimonopole pairs dissolve, and the theory's vacuum becomes a gas of monopoles and antimonopoles without strong correlations. 
In light of our results, we numerically demonstrate that the spectrum of transitional/collective phenomena occurring in $U(1)$-LGT is actually richer than the expected one. Indeed, the pairs dissolution is only one of the mechanisms occurring in an Abelian LGT. 
For this aim, we consider the 3D monopoles (topological defects) of the gauge field
defined according to \cite{degrand1980topological,tobochnik1979monte} (see SM for more details). In our simulations, we compute the average densities of monopole-antimonopole pairs at the nearest-neighbors distance, $\langle\rho_{pair}\rangle$, and of isolated monopoles, $\langle\rho_{mon}\rangle$. 
The yellow and violet curves in Fig.~\ref{fig:derS_3_monopoles} represent $\langle\rho_{pair}\rangle$ and $\langle\rho_{mon}\rangle$, respectively, as functions of the energy.
At $\epsilon\leq \epsilon^{ind}_{3\,(a)}$, monopoles are not 
detected, and $\langle\rho_{mon}\rangle=\langle\rho_{pairs}\rangle=0$.
The first appearance of monopoles, i.e., non-vanishing values for 
$\langle\rho_{mon}\rangle$ and $\langle\rho_{pairs}\rangle$ occurs around 
$\epsilon_{first}\approx 0.62$ (see the inset in Fig.~\ref{fig:derS_3_monopoles}). 
This effect is reflected in $\partial^{3}_{\epsilon}S$ with the presence of a positive-valued minimum of at $\epsilon^{ind}_{3\,(a)}$; this corresponds to an independent third-order PT, according with the more recent version of MIPA \cite{qi2018classification}. 
It should be mentioned that a non-negligible mismatch is found between $\epsilon_{first}$ and $\epsilon^{ind}_{3\,(a)}$.
A discussion about the reasons at the ground of this discrepancy is provided in the SI. 
For energies slightly larger than $\epsilon^{ind}_{3\,(a)}$, only a small number of monopoles is found in bonded pairs; the largest amount of monopoles is in the isolated configuration, i.e., the free states. Increasing the energy, the density of isolated monopoles $\langle\rho_{mon}\rangle$ reaches a local maximum around $\epsilon_{peak}\approx0.78$. Thus, the production of isolated monopoles is favored over that of pairs up to $\epsilon_{peak}$ where such a production ends. For $\epsilon_{peak}\leq\epsilon\leq \epsilon^{ind}_{3\,(b)}$, the process reverses: the isolated monopoles begin to interact \emph{leaving} the isolated configuration to form pairs as one deduces from the subsequent drop of $\langle\rho_{mon}\rangle$. At this stage, the production of pairs increases, reaching a peak about $\epsilon\simeq\epsilon^{ind}_{3\,(b)}$ corresponding to the independent third-order PT.
Therefore, such a transition corresponds to the saturation of the pairs production process; the system reaches a new phase, that is, a new vacuum state of the theory mostly populated by pairs.
Increasing the energy, the density of pairs $\langle\rho_{pairs}\rangle$ decreases crossing over $\epsilon^{ind}_{1}$ and reaching a plateau in correspondence of $\epsilon^{dep}_{3}$. The drop in the density of pairs is the physical process originating from the pair-dissolution mechanism and leading to the DPT.
It is worth noting that, after the pair dissolution (see the vertical red line in Fig.~\ref{fig:derS_3_monopoles}), the slope of $\langle\rho_{pairs}\rangle$ ($\langle\rho_{mon}\rangle$) passes from being negative (positive) to being ideally constant for any $\epsilon>\epsilon^{dep}_{3}$. Such a change corresponds to the dependent PT of third-order associated with a negative-valued maximum in $\partial^{3}_{\epsilon}S$ around $\epsilon^{dep}_{3}$ (see the vertical green line in Fig.~\ref{fig:derS_3_monopoles} and the inset within) signaling the transition of the system to the last vacuum state represented by a gas of monopoles and pairs.

In summary, we have shown that the application of MIPA to a microcanonical formulation of the $U(1)$-LGT allowed us to identify two independent (major) third-order PTs and a dependent (minor) third-order PT, never observed before through canonical ensemble-based approaches.
In practice, microcanonical PTs, corresponding to inflection points of entropy or its derivatives, have been interpreted as qualitative changes in the behavior of topological defects (monopoles) when the  energy is varied.
We found that all the transitional phenomena are driven by the occurrence of topological defects or by their interactions.
In particular, we have observed an independent third-order PT at $\epsilon^{ind}_{3(a)}$ at very low energy associated with the first appearance of monopoles. Increasing the energy, the monopoles interact among them, forming pairs interpreted as bound states of the theory and reaching a saturation point corresponding to a further independent third-order PT around $\epsilon_{3(b)}^{ind}$.
At higher energies, a dissociation-pairs process has been observed in correspondence to the well-known deconfinement transition that is an independent first-order PT according to the MIPA scheme. 
During this transition, we are in the presence of a coexistence of two phases of the theory: pairs (bounded states) and isolated monopoles (free states). At large energy, the ground state of the theory corresponds to a gas of monopoles, and it is signaled by a dependent third-order PT. 
It should be stressed that adopting Kogut's continuum-time formulation of QFT \cite{kogut1983lattice,polonyi1983microcanonical,kogut1987first}, MIPA can be easily extended to systems of fermions interacting with a gauge field as well as to the study of PTs for the Higgs field. 

We conclude by discussing an intriguing similarity between the $U(1)$-LGT, in Villain approximation given by Banks et al.~\cite{banks1977phase}, and many-body systems in condensed matter. Indeed, the Wilson action appearing in the partition function can be recast into an equivalent action where the new degrees of freedom are represented by a monopole density at each lattice site $\{m(\bm{n})\}_{\bm{n}}$. Thus, the action describes a collection of monopoles (many-body system) interacting with each other through the lattice Coulomb potential, i.e., $m(\bm{n})v(\bm{n}-\bm{u})m(\bm{u})$. Since the PTs observed in LGT occur due to the emergence of monopoles and their interactions, by analogy, we could expect new transition phenomena also in condensed-matter systems where the charge density plays a similar role to the monopole density.\\

L.D.C is grateful with Dr. Dario Consoli and Dr. Alfredo Grillo for their helpful suggestions and discussions.

\section*{}
\providecommand{\noopsort}[1]{}\providecommand{\singleletter}[1]{#1}%
%


\newpage
\newpage

\begin{widetext}

\textbf{{\Large SUPPLEMENTAL MATERIAL}}

\renewcommand{\theequation}{S.\arabic{equation}}

\maketitle

\setcounter{equation}{0}

\section{Hamiltonian model for the U(1) compact model on 4D Euclidean lattice: numerical algorithm}

The investigation of the thermodynamic properties of the $U(1)$-LGT is based on the molecular dynamics simulation of the Hamiltonian system defined by
\begin{equation}\label{def:hamiltonian_LGT}
    \mathscr{H}(\pi,\phi):=\sum_{\mu,n}\frac{\pi_{n,\mu}^2}{2}+\sum\limits_{\square_{\mu,n}}\left[1-\cos(\phi_{\bm{n},\mu}+\phi_{n+\mu,\nu}-\phi_{n+\nu,\mu}-\phi_{n,\nu})\right].
\end{equation}
where $\square_{\mu,\nu,n}$ is a representative plaquette. We then numerically solved the Hamilton equations of motion
\begin{equation}
    \frac{d\phi_{\mu,n}}{d\tau}=\pi_{\mu,n},\qquad\frac{d\pi_{\mu,n}}{d\tau}(\tau)=-\frac{\partial\mathscr{H}}{\partial\phi_{\mu,n}},
\end{equation}
adopting a second order bilateral symplectic algorithm \cite{casetti1995efficient} since it conserves the symplectic structure of the Hamiltonian flow. Such an algorithm reads ($\tau_{n/2}:=\tau_0+n\Delta\tau/2$):
\begin{equation}
\begin{split}
    \phi_{\mu,n}(\tau_{1/2}) &= \phi_{\mu,n}(\tau_{0}),\\
    \pi_{\mu,n}(\tau_{1/2}) &= \pi_{\mu,n}(\tau_0)- F_{\mu,n}[\phi(\tau_{1/2})]\Delta \tau/2 ,\\
    \phi_{\mu,n}(\tau_{1})&=\phi_{\mu,n}(\tau_{1/2})+ \pi_{\mu,n}(\tau_{1/2})\Delta\tau,\\
    \pi_{\mu,n}(\tau_{1})&=\pi_{\mu,n}(\tau_{1/2})- F_{\mu,n}[\phi_{1}]\Delta \tau/2,\\
    \pi_{\mu,n}(\tau_{3/2})&=\pi_{\mu,n}(\tau_{1}),\\
    \phi_{\mu,n}(\tau_{3/2})&=\phi_{\mu,n}(\tau_{1})+ \pi_{\mu,n}(\tau_{3/2})\Delta \tau/2,\\
    \pi_{\mu,n}(\tau_{2})&=\pi_{\mu,n}(\tau_3/2)- F_{\mu,n}[\phi(\tau_{3/2})]\Delta \tau,\\
    \phi_{\mu,n}(\tau_{2})&=\phi_{\mu,n}(\tau_{3/2})+ \pi_{\mu,n}(\tau_{3/2})\Delta\tau,
\end{split}
\end{equation}
where $F_{\mu,n}(\tau_{n/2}):=\partial_{\phi_{\mu,n}}\mathscr{V}[\phi(\tau_{n/2})]$. The time step has been chosen to be $\Delta\tau=0.01$ and it leads to a conservation of energy around $10^{-6}$-$10^{-7}$. The initial conditions have been chosen randomly for $\phi_{\mu,n}$ and zero for $\pi_{\mu,n}$ (as explained later). In this way, we obtained a certain energy value, $E_{in}$, and we allowed the system to evolve for $10^{6}$ steps so as to equilibrate the trajectory. After this procedure, if $E_{in}$ was not the desired energy, say $E_{des}$, we reached the energy value $E_{des}$ by choosing a suitable positive (negative) constant $\alpha$ such that, multiplying each momentum $\pi_{\mu,n}$ by $\alpha$, the system was heated (cooled). In practice, starting from $E_{in}$, one guesses a $\alpha_{guess}$ and measures $E_{guess}=\mathscr{H}[\alpha\pi_{\mu,n},\phi_{\mu,n}]$. This procedure is repeated as long as $|E_{guess}-E_{des}|$ is smaller than the chosen precision. Once the desired energy has been reached, the system is equilibrated again for $10^{6}$ iterations before saving the trajectory used for computing averages. Each thermodynamic observable has been evaluated on $N_{step}=131071$ measurements, and the definition of the microcanonical average is based on the ergodic assumption that, adapted to simulations, reads
\begin{equation}\label{def:time_average}
    \langle f\rangle_{\varepsilon}=\frac{1}{N_{step}}\sum_{n=1}^{N_{step}}f(\tau_n).
\end{equation}
The constraint $\pi_{\mu,n}=0$ for the initial values has been proposed by Callaway in Refs. \cite{callaway1982microcanonical,callaway1983lattice} to eliminate the extra degrees of freedom coming from the gauge invariance. This procedure is equivalent to imposing a gauge such as, for instance, the Lorentz gauge or the axial one.

\section{Calculation of Microcanonical Entropy Derivatives}

The Microcanonical Inflection Point Analysis (MIPA) requires to calculate the 
derivatives of the specific microcanonical entropy defined as
\begin{equation}
    S_{N_{\rm ind}}(E) = \log \Omega_{\rm ind}(E)=\log \int \delta(H(\mathbf{\pi},\mathbf{\phi})- E) \delta(\mathscr{C}(\mathbf{\pi})) \,\,\,\mathrm{D}\mathbf{\pi}\mathrm{D}\mathbf{\phi}
\end{equation}
where the condition $\mathscr{C}(\mathbf{\pi})=0$ is equivalent to a gauge-fixing condition (see the previous 
section) and $N_{\rm ind} = 3 N_{\rm lat}^4$ is the number of independent momenta \cite{callaway1982microcanonical,callaway1983lattice}.
The specific microcanonical entropy has been introduced so as
to extract extensive quantities and compare systems of 
different sizes: its form is $s_{N_{\rm ind}}(\varepsilon)=N_{\rm ind}^{-1}S_{N_{\rm ind}}(N_{\rm ind}\varepsilon)$. 

\subsection{First- and second-order derivative of microcanonical entropy}
As shown in Ref. \cite{pearson1985laplace}, in the microcanonical ensemble, given any fixed value of the specific energy, the derivatives of specific entropy can be expressed in terms of the moments of the distribution of the inverse specific kinetic energy $k=K(\boldsymbol{\pi})/N_{\rm ind}$. In this framework,  the first- and second-order derivative of specific microcanonical entropy can be written as 
\begin{equation}
\label{eq:I_II_derMicroCanS}
\begin{split}
    & \partial_{\varepsilon}s_{N_{\rm ind}}(\varepsilon)= \left(\frac{1}{2}-\frac{1}{N_{ind}}\right)\langle k^{-1}\rangle_{\varepsilon} \\
    & \partial^{2}_{\varepsilon}s_{N_{\rm ind}}(\varepsilon)
    =N_{ind}\bigg[\bigg(\dfrac{1}{2}-\dfrac{1}{N_{ind}}\bigg) \bigg(\dfrac{1}{2}-\dfrac{2}{N_{ind}}\bigg)\langle k^{-2} \rangle_{\varepsilon}
    - \bigg(\dfrac{1}{2}-\dfrac{1}{N_{ind}}\bigg)^2 \langle k^{-1}\rangle^2_{\varepsilon} \bigg]\\
\end{split}
\end{equation}
where the microcanonical averages have been estimated by Eq.\eqref{def:time_average}.
It is worth noting that the standard thermodynamic observables as the 
microcanonical temperature, $T(\varepsilon)$, and the microcanonical 
specific heat, $C_v (\varepsilon)$, given by
\begin{equation}
    \begin{split}
        & T_{\mu c}(\varepsilon)=\beta_{\mu c}=(\partial s/\partial \varepsilon)^{-1}\\
        &C_v (\varepsilon)=(\partial T(\varepsilon)/\partial \varepsilon)^{-1}\equiv -(\partial s/\partial \varepsilon)^{2}(\partial^{2} s/\partial \varepsilon^{2})^{-1}\,\,.
    \end{split}
\end{equation}
are connected to the derivatives of microcanonical entropy and, thereby, they can be computed by means of Eqs.~\eqref{eq:I_II_derMicroCanS}.
\begin{figure*}[h!]
    \includegraphics[scale=0.275]{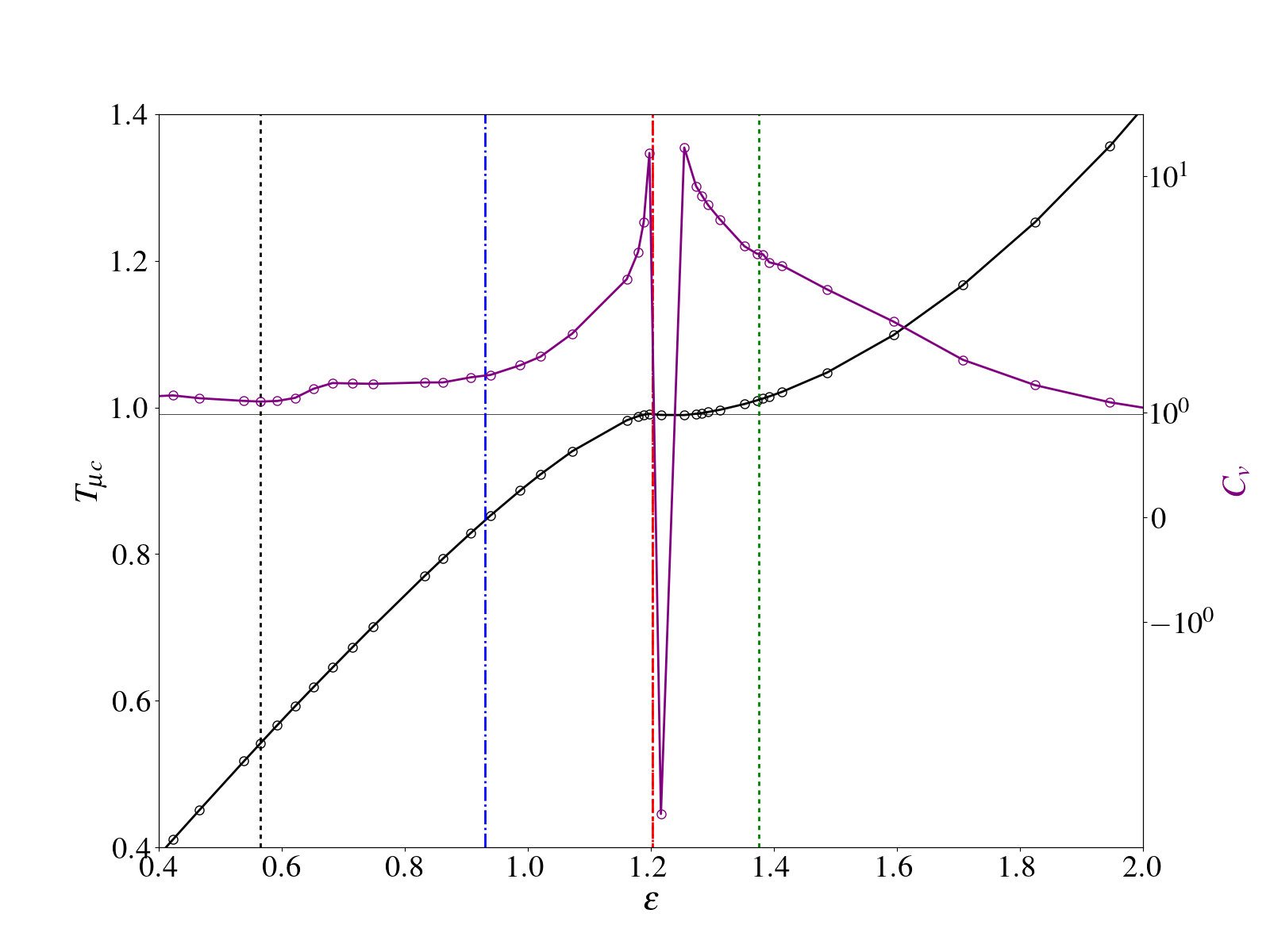}
    \caption{\textbf{Microcanonical thermodynamic observables in U(1) LGT}. The microcanonical temperature $T_{\mu c}$ (black circles) and the microcanonical specific heat $C_v $ (purple circles) vs. the specific energy $\varepsilon$ have been reported for the $U(1)$-LGT with a cubic lattice with $N=12$ sites for side.}
\end{figure*}

\subsection{Third-order derivative of microcanonical entropy}

In principle, the third-order derivative of microcanonical entropy can be estimated using the same formalism presented in the previous subsection.\\
However, the poor convergence of the third-order moment of kinetic energy distribution does not allow calculating the third-order derivative of microcanonical entropy with such a method. Thus numerical derivative of the second-order derivative $\partial_{\varepsilon}^2 s$ of specific microcanonical entropy  has been estimated.\\

\section{On the origin of the S-shape in the plaquette potential vs microcanonical entropy}

We report in this section the behavior of the plaquette 
potential $\langle \mathscr{V}_{\square} \rangle_{\mu c}$
and of its energy derivative $\partial_{\varepsilon}\langle \mathscr{V}_{\square} \rangle_{\mu c}$ regarded as functions of 
specific energy (see Fig.~\ref{fig:potVvsEn} below). Observing such behaviors, we can conclude that in a neighborhood 
of the deconfinement phase transition, the energy 
derivative of the average plaquette potential is bounded.

\begin{figure*}[h!]
    \includegraphics[scale=0.335]{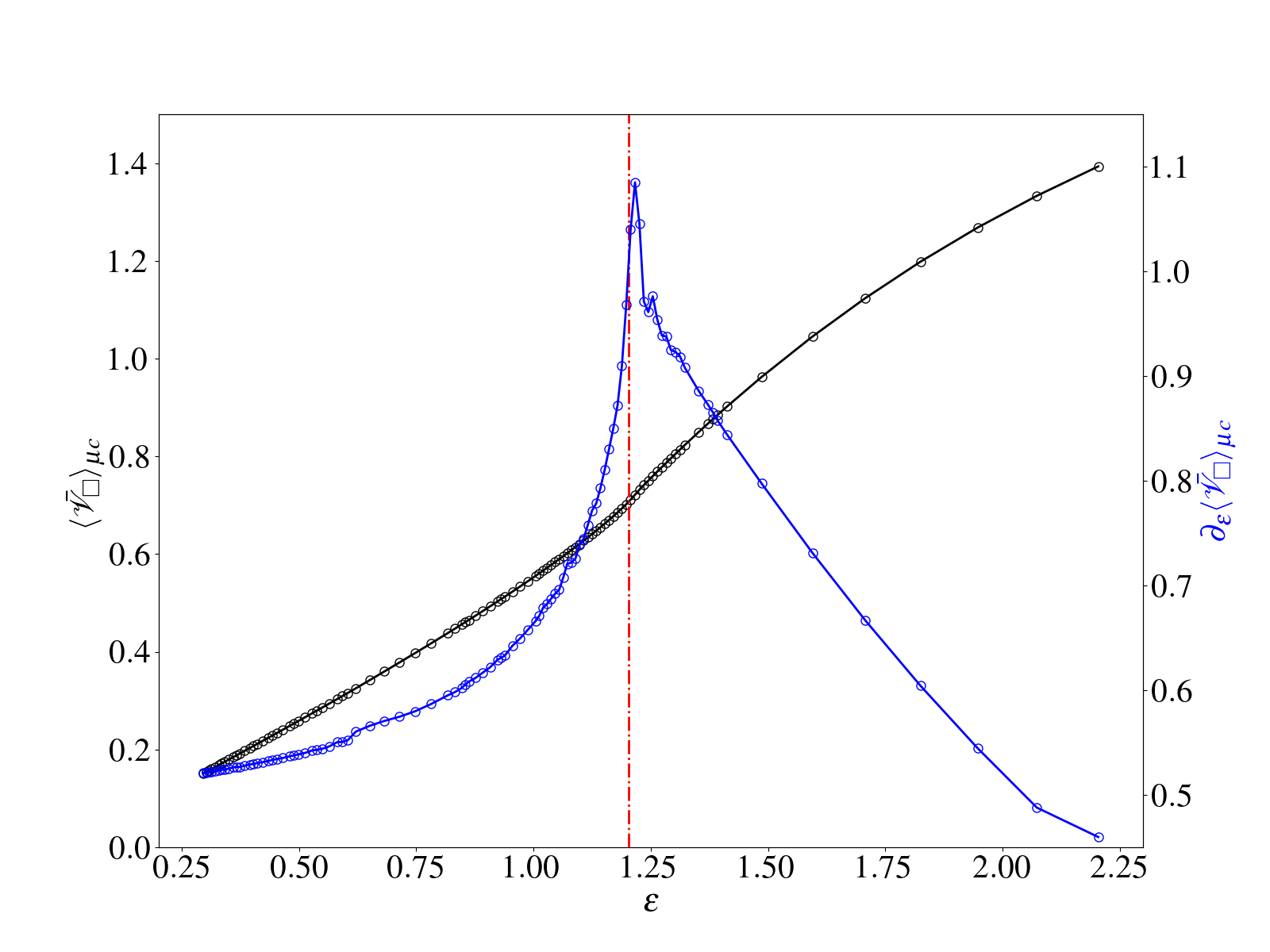}
    \caption{\textbf{Microcanonical plaquette potential vs specific energy}. The microcanonical average of the plaquette potential (black line) and its derivative (blue line) are plotted as functions of the specific energy. The dashed-dotted red line indicates the critical energy, $\epsilon^{ind}_{1}$, for the deconfinement transitions}
    \label{fig:potVvsEn}
\end{figure*}




\section{Computation of monopoles}

The computation of the topological defects used in our work is based on Ref.~\cite{degrand1980topological}. In particular, we studied 3D monopoles redefining the 3D plaquette angle as $\bar{\Theta}_{ij}=\Theta_{ij}-2\pi K_{ij}$ where $K_{ij}:=\text{mod}(\Theta_{ij},2\pi)$. Then, using Gauss' law, we get:
\begin{equation}\label{def:number_monopoles}
    N(\bm{n}) = \sum_{i,j,k=1}^{3} \epsilon_{ijk}(K_{jk}(\bm{n}+\bm{e}_{i})-K_{jk}(\bm{n})).
\end{equation}
Here, $\epsilon_{ijk}$ is the Levi-Civita antisymmetric tensor. By definition, $\phi_{\bm{n},\mu}\in[-\pi,\pi]$ and $\Theta_{ij}\in[-4\pi,4\pi]$, then, $N=0,\,\pm 1,\,\pm 2$. If $N=0$ no topological defects are present in the plaquette defined by the site $\bm{n}$ and the three directions $\bm{e}_{x}$, $\bm{e}_{y}$ and $\bm{e}_{z}$ whereas if ($N<0$) $N>0$ we are in presence of (anti)monopoles with charge 1 or 2. We study the average density of isolated monopoles $\langle\rho_{mon}\rangle$ and the total number of monopole-antimonopole pairs $\langle\rho_{pairs}\rangle$. The calculation of the former is summarized in the following procedure. At any lattice site, $\bm{n}$, we identify $N_{iso}(\bm{n})$ to be the number computed through Eq.~\eqref{def:number_monopoles} such that \textit{(i)} $N(\bm{n})\neq 0$, \textit{(ii)} $N(\bm{n}\pm\bm{e}_{x})=N(\bm{n}\pm\bm{e}_{y})=N(\bm{n}\pm\bm{e}_{y})=0$. Proceeding in this manner for each lattice site, we define
\begin{equation}\label{def:density_isolated}
    \langle\rho_{mon}\rangle=\frac{1}{V}\sum_{\bm{n}}N_{iso}(\bm{n}),
\end{equation}
where $V=N^4$ is the volume of the lattice. In our case, $N=12$. For what concerns the number density of pairs, we proceed similarly as before but we identify $N_{pairs}(\bm{n})$ with the number computed in Eq.~\eqref{def:number_monopoles} if \textit{(i)} $N(\bm{n})\neq0$, \textit{(ii)} $N(\bm{n}+\bm{e}_{i})=-N(\bm{n})$ only for a single $i=x,\,y,\,z$. Thus, the associated observable is
\begin{equation}\label{def:density_pairs}
    \langle\rho_{pairs}\rangle=\frac{1}{V}\sum_{\bm{n}}N_{pairs}(\bm{n}).
\end{equation}

\section{Apparent discrepancy for the first pairs-appearance phase transition}

As mentioned in the main text, we associated the independent third-order PT occurring at $\epsilon^{ind}_{3(a)}\approx 0.57$ to the first appearance of pairs and monopoles. This conceptual link is suggested by the presence of a positive-valued minimum in the third-order derivative of entropy and by non-vanishing values of the number density of pairs $\langle\rho_{pairs}\rangle$ around $\epsilon_{first}\approx0.62$. Nevertheless, as one can see in Fig.~3, these two manifestations do not occur exactly at the same energy and a mismatch is found $|\epsilon_{first}-\epsilon_{3\,(a)}^{ind}|\approx0.05$. We believe that the presence of this mismatch has actually a numerical origin due to two possible reason: large average waiting time for the occurrence of monopoles and the definition of monopoles. For what concerns the first, we notice that the estimation of the energy associated to the first appearance of topological defects can be refined by means of longer simulations at low energies. Indeed, we note that at low energies the system is ``frozen''; roughly speaking, the representative point of the trajectory evolves very slowly from its initial configuration. Thus, we can image that the average waiting time, $\tau_{wait}(E)$, for the occurrence of a monopole which depends on energy decreases as the energy increases. In practice, in our simulation we fixed the time length of the trajectories to be $T_{sim}=N_{step}\cdot\Delta t=1310.71$. At very low energies, $E\approx 0$, we have that $T_{sim}/\tau_{wait}(E)\approx 0$. Hence, we should extend the simulations, in principle, with $T_{sim}\to\infty$. Increasing the energy, we found that $\tau_{wait}(E)$ becomes smaller then $T_{sim}$ only from energies $\epsilon\geq\epsilon_{first}\approx0.62$ where the first appearance of monopoles has been detected. In conclusion, we expect that for $T_{sim}$ larger than that used now monopoles can be detected earlier. On the other hand, we also expect that there exists a certain threshold indicated by the thermodynamic observable below which monopoles are detected only after an infinite average waiting time. Regarding the second reason that we mentioned above, we notice that the definition of monopoles is based on the integers $K_{ij}=\text{mod}(\Theta_{ij},2\pi)$ where $\Theta_{ij}$ is the plaquette angle that is a real number. Thus, for any value of $\Theta_{ij}$ that is not greater or exactly equal to $2\pi$, $K_{ij}$ will be always zero. In other words, Eq.~\eqref{def:number_monopoles} does not recognize any monopole arising from those plaquette angles such that $\Theta_{ij}=2\pi-\delta$ with an arbitrary small $\delta>0$. In contrast to that, the calculation of kinetic energy used for computing the thermodynamics observables is not subjected to any constraint expected those given by the equations of motion. Thus, the signals in the thermodynamic observable can appear at the correct energy value.

\end{widetext}

\end{document}